\documentclass[12pt]{article}
\pdfoutput=1
\usepackage{latexsym, graphicx,cite,mathrsfs}
\usepackage{amsmath}
\usepackage{amssymb}
\usepackage{amsthm}

\allowdisplaybreaks

\usepackage[top = 1. in,bottom = 1 in, left = 1 in, right=1 in]{geometry}
\usepackage[colorlinks=true, linkcolor=blue, bookmarks=true]{hyperref}
\newcommand{\arXiv}[1]{\href{http://www.arXiv.org/abs/#1}{arXiv:#1}}
\newcommand{\doi}[2]{\href{http://dx.doi.org/#2}{#1}}

\makeatletter
\renewcommand\section{\@startsection {section}{1}{\z@}%
                  {-3.5ex \@plus -1ex \@minus -.2ex}
                  {2.3ex \@plus.2ex}%
                  {\normalfont\large\bfseries}}
\renewcommand\subsection{\@startsection{subsection}{2}{\z@}%
                   {-3.25ex\@plus -1ex \@minus -.2ex}%
                   {1.5ex \@plus .2ex}%
                   {\normalfont\bfseries}}
\makeatother

\newcommand{\beq}{\begin{equation}}
\newcommand{\eeq}{\end{equation}}
\newcommand{\ber}{\begin{array}}
\newcommand{\eer}{\end{array}}
\newcommand{\del}{\partial}

\newcommand{\ph}{\varphi}
\newcommand{\de}{\delta}
\newcommand{\al}{\alpha}
\newcommand{\be}{\beta}
\newcommand{\ga}{\gamma}
\newcommand{\la}{\lambda}
\newcommand{\ze}{\zeta}
\newcommand{\bea}{\begin{eqnarray}}
\newcommand{\eea}{\end{eqnarray}}
\newcommand{\ER}{Erd\H{o}s-R\'enyi }
\renewcommand{\Re}{\operatorname{Re}}
\renewcommand{\Im}{\operatorname{Im}}
\newcommand{\Tr}{\mathrm{Tr}}

\newcommand{\dagg}{\dagger}
\newcommand{\nn}{\nonumber}
\newcommand{\floor}[1]{\lfloor #1 \rfloor}

\begin{document}
\begin{titlepage}
\begin{flushright}
\phantom{start}
\end{flushright}
\vspace{5mm}
\begin{center}
{\LARGE\bf Statistical field theory of random graphs\vspace{2mm}\\ with prescribed degrees}\\
\vskip 12mm
{\large Pawat Akara-pipattana$^{a}$ and Oleg Evnin$^{b,c}$}
\vskip 10mm
{\em $^a$ Universit\'e Paris-Saclay, CNRS, LPTMS, 91405 Orsay, France}
\vskip 3mm
{\em $^b$ High Energy Physics Research Unit, Department of Physics, Faculty of Science, Chulalongkorn University,
10330 Bangkok, Thailand}
\vskip 3mm
{\em $^c$ Theoretische Natuurkunde, Vrije Universiteit Brussel and\\
 International Solvay Institutes, 1050 Brussels, Belgium}
\vskip 7mm
{\small\noindent {\tt pawat.akarapipattana@universite-paris-saclay.fr, oleg.evnin@gmail.com}}
\vskip 20mm
\end{center}
\begin{center}
{\bf ABSTRACT}\vspace{3mm}
\end{center}
Statistical field theory methods have been very successful with a number of random graph and random matrix problems, but it is challenging
to apply these methods to graphs with prescribed degree sequences due to the extensive number of constraints that enforce the desired degree at each vertex.
Building on top of recent results where similar methods are applied to random regular graph counting, we develop an accurate statistical field theory description for the adjacency matrix spectra of graphs with prescribed degrees. For large graphs,
the expectation values are dominated by functional saddle points satisfying explicit equations. For the case of regular graphs, this immediately
leads to the known McKay distribution. We then consider mixed-regular graphs with $N_1$ vertices of degree $d_1$,  $N_2$ vertices of degree $d_2$, etc, such that the ratios $N_i/N$ are kept fixed as $N$ goes to infinity. For such graphs, the eigenvalue densities are governed by nonlinear integral equations of the Hammerstein type. Solving these equations numerically reproduces with an excellent accuracy the empirical eigenvalue distributions.

\vfill

\end{titlepage}

\section{Introduction}

Statistical field theory methods \cite{statFT,statFTneu} have found fruitful applications to a number of problems involving random matrices and random graphs; see \cite{BR,RB,FM,MF,SC,PN,Guhr,euclRM,kuehn,spectrum,AC,TO,resdist,RP2022,laplace,Baron,walkongraph} for a sampler of literature.
When applied to the topic of spectra of random matrices, as in \cite{BR,RB,FM,MF,Guhr,euclRM,kuehn,spectrum,RP2022,laplace}, this program usually  starts with the resolvent, or the Stieltjes transform of the eigenvalue density $p(\la)$ in the mathematical language,
\beq
R(z)=\int_{-\infty}^\infty\hspace{-2mm} d\la\,\frac{p(\la)}{\la-z},
\eeq
from which the eigenvalue density can be recovered via the Sokhotski-Plemelj formula as
\beq
p(\la)=\frac1\pi\Im\hspace{-0.3mm} R(z)\Big|_{z=\la+i0}=-\frac1\pi\Im\hspace{-0.3mm} R(z)\Big|_{z=\la-i0}.
\eeq
The resolvent can be expressed through matrix inverses which, in turn, have nice representations in terms of Gaussian integrals that lead to factorization over the entries of the matrices, so that the random matrix averaging is straightforwardly performed. One is left with a vector model that is solvable at large $N$, where $N$ is the original matrix size. To make this procedure go through, one needs to ensure that the determinant factors naturally arising from the Gaussian integrals at the intermediate stages all drop out, and one nice way to accomplish this is via introducing integrals over anticommuting variables, leading to `supervector' models. Textbook introductions to these topics can be found in \cite{efetov,wegner}, some early influential applications of relevance for us here are in \cite{MF,FM}, and we have provided a pedagogical introduction to this technology in \cite{laplace}.

When dealing with regular graphs (graphs with vertices that are all of the same prescribed degree), and more generally, with random graphs that have different prescribed degrees for different vertices, this program meets substantial difficulties. The reason is that there is an extensive number of hard constraints (a degree specification at each vertex), and inserting this huge number of constraints into statistical field theory averages induces considerable complications. Physically, the presence
of an extensive number of constraints is known to lead to strong consequences from the standpoint of thermodynamics, see for instance \cite{breaking}. Random graphs with prescribed degree sequences have been considered from the mathematical perspective in \cite{degseq}.
Treatments of spectral densities for graphs with prescribed degrees have appeared in the math \cite{empi} and physics \cite{SM} literature, but with a focus on high degrees. The target here is to develop a theory of spectral densities for finite degrees (independent of the graph size), which is more challenging. Degrees in physical networks are typically determined by the local properties of the nodes, and do not grow
with the system size.

Our goal in this article is to overcome the above difficulties and present a successful treatment for eigenvalue distributions of graphs with prescribed degrees
based on statistical field theory methods. We essentially build upon the recent considerations of the simpler problem of counting graphs with prescribed degrees in \cite{Kawamoto,regcount} that rely on compatible methods. In particular, in \cite{regcount} a pattern has been identified in the saddle point structure that makes the problem analytically manageable, and it will be directly reused in our treatment below.

After these generalities have been laid out in sections~\ref{secaux} and \ref{secsddl}, we shall first test our formalism in section~\ref{secMcKay} by applying it to random regular graphs, where it straightforwardly reproduces the known McKay distribution \cite{mckay}. We then turn in section~\ref{secmix} to mixed-regular graphs, that is, graphs on $N$ vertices where the first $N_1$ vertices are of degree $d_1$,  the subsequent $N_2$ vertices of degree $d_2$, and so on, such that the ratios $N_i/N$ are kept fixed as $N$ goes to infinity. In this case, there are no explicit analytic formulas for the eigenvalue densities, but our theory produces a saddle point problem in the form of a nonlinear integral equation of the Hammerstein type \cite{surv}. These equations have Bessel kernels, which are also commonly seen in other related sparse random matrix problems
\cite{BR,RB,MF,FM,laplace,walkongraph,gel,Khetal,inhmg}. We have recently spelled out a simple and effective numerical strategy for handling such equations
\cite{hmmr}, and we shall employ a variation of this strategy to solve the equations that arise for the case of mixed-regular graphs. The numerical results
will be seen to accurately reproduce the empirically observed eigenvalue distributions. We conclude with a summary and discussion in section~\ref{secconc}.

\section{Auxiliary field representation}\label{secaux}

For treatments of random matrices in the style of statistical physics, it is a well-established and effective approach to represent the eigenvalue density $p(\la)$ of an $N\times N$ matrix $\mathbf{A}$ as
\beq\label{rhodef}
p(\la)\equiv\frac1{N} \sum_{k=1}^N\de(\la-\la_k)=\frac1{\pi N}\Im\sum_k\frac1{\la-\la_k-i0}=-\frac1{\pi N}\Im\Tr[\mathbf{A}-z\mathbf{I}]^{-1}\Big|_{z=\la-i0},
\eeq
where $\la_k$ are the eigenvalues of $\mathbf{A}$, and we have used the Sokhotski–Plemelj formula together with the spectral decomposition of the matrix inverse. The notation $z=\la-i0$ implies evaluating the relevant expression just below the real axis. The resolvent $\Tr[\mathbf{A}-z\mathbf{I}]^{-1}$ is
a rational function of the matrix entries, an improvement over the individual eigenvalues that are not expressible as simple functions of the entries of $\mathbf{A}$. 

One then represents the resolvent in terms of Gaussian integrals. To do it effectively, we introduce \cite{MF,FM,efetov,wegner} a 4-component `supervector' at every vertex $k$, given by $\Psi_k\equiv (\phi_k,\chi_k,\xi_k,\eta_k)$. Here, $\phi_k$ and $\chi_k$ are ordinary (bosonic) numbers, while $\xi_k$ and $\eta_k$ are anticommuting (fermionic) numbers satisfying
\beq\label{Berezin}
\xi_i\xi_j=-\xi_j\xi_i,\qquad \xi_i^2=0,\qquad \int d\xi_i \xi_j=\de_{ij},\qquad \int d\xi_i=0.
\eeq
The components of $\xi$ also anticommute with the components of $\eta$. The `Berezin integration rules' given above are a convenient shorthand
that generates attractive algebraic properties in the resulting expressions, for example, the following important Gaussian integration formula valid for any matrix $\mathbf{M}$:
\beq\label{Grassdet}
\int d\xi_i\,d\eta_j\,e^{-\sum_{kl}M_{kl}\xi_k\eta_l}=\det\mathbf{M}.
\eeq
We define a `scalar product' of supervectors as
\beq\label{scalarp}
\Psi^\dagg_k\Psi_l\equiv \phi_k\phi_l + \chi_k\chi_l +\frac{\xi_k\eta_l-\eta_k\xi_l}{2},\qquad (\Psi^\dagg_k\Psi_k\equiv \phi_k^2 + \chi_k^2 +\xi_k\eta_k).
\eeq
Then, with (\ref{Grassdet}) and the standard Gaussian integrals for bosonic variables, we can write
\begin{equation}\label{Gauss}
\frac1N\Tr[\mathbf{A} - z\mathbf{I}]^{-1}\equiv\frac1{N} \sum_k[\mathbf{A} - z\mathbf{I}]^{-1}_{kk} = \int d\mathbf{\Psi}\, \Big(\frac{2i}{N}\sum_k\phi_k^2\Big) \,e^{i\sum_{kl}A_{kl}\Psi^\dagg_k\Psi_l - iz\sum_k \Psi^\dagg_k\Psi_k}.
\end{equation}
The purpose of introducing anticommuting variables and Berezin integrals in (\ref{Gauss}) is to cancel the powers of determinants that would have otherwise emerged from standard Gaussian integrals. With the determinants cancelled out, the representation is well-adapted for further averaging over $\mathbf{A}$. The presentation of (\ref{Gauss}) we have given here is completely standard in the literature on supersymmetric statistical field theory methods in application to random matrix problems, and for that reason we did not dwell much on any details. We have previously provided a pedagogical introduction to these techniques in \cite{laplace}, and advise the readers less familiar with this material to consult the opening sections of that paper.
Mathematically rigorous approaches to similar representations have been explored in \cite{shch1,shch2}.

Equation (\ref{Gauss}) as it stands is valid for any prescribed matrix $\mathbf{A}$. If we were to average it over an ensemble where the entries of $\mathbf{A}$ are independent, the factorization of the right-hand side of (\ref{Gauss}) over the entries of $\mathbf{A}$ would make the averaging straightforward, leaving behind a large $N$ supervector model in terms of $\pmb{\Psi}$. The situation is more intricate 
when $\mathbf{A}$ is the adjacency matrix for a graph with prescribed degrees, which is what interests us here. Recall that the entries $A_{ij}=A_{ji}$ of the adjacency matrix
equal 1 if there is an edge between vertices $i$ and $j$, and 0 otherwise. If vertex $k$ is assigned degree $d_k$, the entries of $\mathbf{A}$
must satisfy $N$ constraints:
\beq\label{degs}
\sum_{l=1}^N A_{kl}=d_k.
\eeq
Handling the statistical field theory representation for the eigenvalue distributions of adjacency matrices in the presence of such constraints will
be the main technical content of this paper.

The rest of our treatment will be a `marriage' between the considerations of \cite{laplace}, whose origins can be further traced back to \cite{MF,FM}, 
and the techniques developed in \cite{regcount} for the purpose of regular graph counting. From \cite{laplace}, we inherit the Fyodorov-Mirlin method \cite{MF,FM} for processing in the large $N$ limit the supervector model emerging from (\ref{Gauss}). From \cite{regcount}, we inherit the strategy for handling the
constraints (\ref{degs}) within a statistical field theory setting, and the nontrivial identification of the saddle point structure that dominates the final result.

An unconstrained summation over all possible adjacency matrices can be implemented by expressing the summand through the independent entries of the symmetric matrix $\mathbf{A}$, that is, $A_{kl}$ with $k<l$, and then summing over the values of 0 and 1 for each of these independent entries.
To implement the degree constraints (\ref{degs}), we can insert into this sum a function that equals 1 if (\ref{degs}) is satisfied and 0 otherwise. To construct such a function, we start with the evident observation that, for any integer $n$,
\beq
\frac{1}{2\pi i}\oint\frac{d\ze}{\ze^{n+1}}=\begin{cases}1
&\mbox{if}\quad  n=0\\0&\mathrm{otherwise,}
\end{cases}
\eeq
where the complex plane integration contour encircles the origin. Using the difference of the right-hand side and left-hand side of (\ref{degs}) in place of $n$, for each $k$, gives us for the average of the resolvent over graphs with the given degree sequence $d_k$
\beq\label{resav}
\frac1N\langle\Tr[\mathbf{A} - z\mathbf{I}]^{-1}\rangle 
=\frac1{\mathcal{N}N} \sum_{\mathbf{A}} \oint \prod_k\left(\frac{d\ze_k}{2\pi i \ze_k^{d_k+1}}\,\ze_k^{\sum_l A_{kl}}\right)
\Tr[\mathbf{A} - z\mathbf{I}]^{-1}.
\eeq
where the sum is over all adjacency matrices $\mathbf{A}$, and $\ze_k$ with $k=1..N$ are complex variables, each of them integrated over a contour that encircles the origin. The normalization factor $\mathcal{N}$ is the total number of graphs with the given degree sequence. We will not need to analyze this normalization factor in any detail as it will drop out in the final result. Expressing (\ref{resav}) through the independent entries $A_{kl}$ with $k<l$, we get
\beq
\frac1N\langle\Tr[\mathbf{A} - z\mathbf{I}]^{-1}\rangle =\frac1{\mathcal{N}N} \sum_{\mathbf{A}} \oint \left(\prod_k\frac{d\ze_k}{2\pi i \ze_k^{d_k+1}}\right) \prod_{k<l}(\ze_k\ze_l)^{A_{kl}}\,\Tr[\mathbf{A} - z\mathbf{I}]^{-1}.
\eeq
In combination with the supervector Gaussian integral representation for the resolvent given by (\ref{Gauss}), this yields
\begin{align}
\frac1N\langle\Tr[\mathbf{A} - z\mathbf{I}]^{-1}\rangle  
=\sum_{\mathbf{A}} \int d\pmb{\Psi}\,&\Big(\frac{2i}{\mathcal{N}N}\sum_k\phi_k^2\Big) \,\oint \prod_k\left(\frac{d\ze_k}{2\pi i \ze_k^{d_k+1}}\right) \\
&\times\left[\prod_{k<l}(\ze_k\ze_l)^{A_{kl}} e^{2  iA_{kl} \Psi^\dagger_k \Psi_l}\right]e^{-iz\sum_k \Psi^\dagger_k\Psi_k}.\nn
\end{align}
Due to factorization over the independent entries of $\mathbf{A}$, the summation over $\mathbf{A}$ is performed straightforwardly to yield
\begin{align}
&\frac1N\langle\Tr[\mathbf{A} - z\mathbf{I}]^{-1}\rangle\! =\!\int\! d\pmb{\Psi}\,\Big(\frac{2i}{\mathcal{N}N}\sum_k\phi_k^2\Big)\!\oint\! \left(\prod_k\frac{d\ze_k}{2\pi i \ze_k^{d_k+1}}\right) e^{-iz\sum_k \Psi^\dagger_k\Psi_k}\prod_{k<l}\left(1+\ze_k\ze_le^{2  i \Psi^\dagger_k \Psi_l}\right)\nn\\
&=\int \!d\pmb{\Psi}\,\Big(\frac{2i}{\mathcal{N}N}\sum_k\phi_k^2\Big)\!\oint\! \left(\prod_k\frac{d\ze_k}{2\pi i \ze_k^{d_k+1}}\right)e^{-iz\sum_k \Psi^\dagger_k\Psi_k}\exp\left[ \sum_{k<l}\log\left(1+\ze_k\ze_le^{2  i \Psi^\dagger_k \Psi_l}\right)\right].\label{explog}
\end{align}
To orient ourselves, we can focus on the $\ze_k$ integral and set $\Psi_k=0$ for a second. In that case, we obtain exactly the same structure that 
appeared in \cite{regcount} for the problem of counting regular graphs. Guided by the considerations there, as well as \cite{Kawamoto}, we can expand the logarithm in the last line as
\beq
\log(1+x)=x-\frac{x^2}2+\frac{x^3}{3}-\frac{x^4}{4}+\cdots
\label{logexp}
\eeq
The infinite series may seem worrisome, but if we assume that the degree sequence is bounded from above by $D$, that is,
\beq\label{maxdeg}
d_k\le D,
\eeq
we can truncate the expansion (\ref{logexp}) {\it exactly} at the first $D$ terms. This is because any power of $\ze_k$ higher than $D$
coming from the expansion of the exponential in (\ref{explog}) will be automatically annihilated by the complex contour integrals under the assumption of (\ref{maxdeg}).
The output of these manipulations is
\begin{align}\label{beforeMF}
\frac1N\langle\Tr[\mathbf{A} - z\mathbf{I}]^{-1}\rangle =&\int d\pmb{\Psi}\,\Big(\frac{2i}{\mathcal{N}N}\sum_k\phi_k^2\Big)\,\oint \left(\prod_k\frac{d\ze_k}{2\pi i \ze_k^{d_k+1}}\right)e^{-iz\sum_k \Psi^\dagger_k\Psi_k}\\
&\hspace{2cm}\times\exp\left[-\sum_{k<l}\sum_{J=1}^{D} \frac{(-1)^J (\ze_k\ze_l)^Je^{2  i J \Psi^\dagger_k \Psi_l}}J\right].\nn
\end{align}
At this point, we are left with a (super)vector model of $N$ variables $\Psi_k$ and $N$ variables $\ze_k$, which, most generally, is expected to be 
rather tractable at large $N$. The problem is that the second line in (\ref{beforeMF}) couples the variables in pairs. Fyodorov and Mirlin proposed
in \cite{FM,MF} a general method for decoupling such dependences resulting in factorization over sites. The price to pay is the introduction of Gaussian
functional integrals, but the resulting structure is very manageable.

In the most basic implementation of the Fyodorov-Mirlin method, as reviewed in \cite{laplace}, one introduces a number-valued function $g(\Psi)$ of the supervector $\Psi$ and writes the following Gaussian integration formula:
\beq
\int \mathcal{D}g \exp\left[-\frac{N}2\int d\Psi\,d\Psi'\, g(\Psi) \,C^{-1}(\Psi,\Psi')\,g(\Psi')+i\sum_k g(\Psi_k)\right]=e^{-\frac1{2N}\sum_{kl}C( \Psi_k, \Psi_l)},
\eeq
where $C$ is arbitrary at this point, and we defined $C^{-1}$ as the inverse of $C$ with respect to the integral convolution: 
\beq
\int d\Psi\, C(\Psi_1,\Psi)\,C^{-1}(\Psi,\Psi_2)=\de(\Psi_1-\Psi_2).
\eeq
We will not need the explicit form of $C^{-1}$ in our manipulations.

To apply this general idea for processing (\ref{beforeMF}), we introduce the following collection of functions 
\beq\label{Cdef}
C_J(\Psi,\Psi')\equiv e^{2iJ \Psi^\dagger \Psi'} , 
\eeq
and correspondingly a collection of functions $g_J(\Psi)$ with $J=1..D$ for functional integration. We then write, for each $J=1..D$: 
\begin{align}
\int \mathcal{D}g_J &\exp\left[-\frac{1}{2J}\int d\Psi\,d\Psi'\, g_J(\Psi) \,C_J^{-1}(\Psi,\Psi')\,g_J(\Psi')+\frac{i^{J+1}}{J}\sum_k \ze_k^J g_J(\Psi_k)\right]\nn\\
&=\exp\left[-\frac{(-1)^J}{2J}\sum_{kl}\ze_k^J\ze_l^J C_J(\Psi_k,\Psi_l)\right]\label{MFtransfrm}\\
&=\exp\left[-\frac{(-1)^J}{J}\sum_{k<l}\ze_k^J\ze_l^J C_J(\Psi_k,\Psi_l)-\frac{(-1)^J}{2J}\sum_k \ze^{2J}_k C_J(\Psi_k,\Psi_k)\right].\nn
\end{align}
If we now use (\ref{MFtransfrm}) to re-express the last line of (\ref{beforeMF}), the dependence on $\Psi_k$ factorizes over $k$ as
\begin{align}
\frac1N\langle\Tr[\mathbf{A} - z\mathbf{I}]^{-1}\rangle =&\int \left(\prod_{J=1}^D\mathcal{D}g_J\right)\exp\left[-\sum_{J=1}^D\frac{1}{2J}\int d\Psi\,d\Psi'\, g_J(\Psi) \,C_J^{-1}(\Psi,\Psi')\,g_J(\Psi')\right]\nn\\
&\hspace{5mm}\times\int \left(\prod_k d\Psi_k\,e^{-iz\Psi^\dagger_k\Psi_k}\right)\,\left(\frac{2i}{\mathcal{N}N}\sum_k\phi_k^2\right)\,\oint\left( \prod_k\frac{d\ze_k}{2\pi i \ze_k^{d_k+1}}\right)\nn\\
&\hspace{1cm}\times\prod_{k=1}^N\exp\left[\sum_{J=1}^D\frac{i^{J+1}}{J} \ze_k^J g_J(\Psi_k)+\sum_{J=1}^{\floor{D/2}}\frac{(-1)^J}{2J} \ze^{2J}_k C_J(\Psi_k,\Psi_k)\right]\label{afterMF}
\end{align}
and the $\Psi_k$ integrals can be evaluated independently. To write the result compactly, 
we introduce the following family of polynomials:
\beq\label{Pdef}
P_d(\al_1,\ldots,\al_d;\be_1,\ldots,\be_{\floor{d/2}})\equiv\oint \frac{d\ze}{2\pi i \ze^{d+1}}\exp\left[\sum_{J=1}^d\frac{i^{J+1}}{J}\ze^J\al_J+\sum_{J=1}^{\floor{d/2}}\frac{(-1)^J}{2J}\ze^{2J}\be_J\right].
\eeq
With $\be_J=1$, these polynomials have already appeared in the analysis of regular graph counting in \cite{regcount}.
They can be expressed through the cycle index polynomials of the symmetric group $S_d$, or through the complete exponential Bell polynomials \cite{Comtet}. In practice, for any given $d$, the polynomial is straightforwardly evaluated using the residues at $\ze=0$. Because of the structure of differentiations involved in the residue computations, each monomial of the form $\prod_k \al_k^{p_k}\be_k^{q_k}$ in $P_d$ must satisfy $\sum_k k(p_k+2q_k)=d$.
Further remarkable simplifications will occur as we go ahead, and the final answer for the eigenvalue density at $N\to\infty$ will only be sensitive
to the term in $P_d$ of the highest degree with respect to $\al_1$, which can be straightforwardly extracted in full generality, and not to any other details.
With all of these preliminaries, we convert (\ref{afterMF}) into
\begin{align}\label{resaux}
&\frac1N\langle\Tr[\mathbf{A} - z\mathbf{I}]^{-1}\rangle  =\frac{1}{\mathcal{N}N} \int \left(\prod_{J=1}^D\mathcal{D}g_J\right)e^{-S[g_1,\ldots,g_D]}\\
&\hspace{2cm}\times\sum_{k=1}^N\frac{\int d\Psi (2i\phi^2) e^{-iz\Psi^\dagger\Psi} P_{d_k}(g_1(\Psi),\ldots,g_{d_k}(\Psi); e^{2i \Psi^\dagger \Psi},e^{4i \Psi^\dagger \Psi},\ldots)}{\int d\Psi  e^{-iz\Psi^\dagger\Psi} P_{d_k}(g_1(\Psi),\ldots,g_{d_k}(\Psi); e^{2i \Psi^\dagger \Psi},e^{4i \Psi^\dagger \Psi},\ldots)},\nn
\end{align}
with the `effective action'
\begin{align}\label{Seff}
&S[g_1,\ldots,g_D]\equiv-\sum_{J=1}^D\frac{1}{2J}\int d\Psi\,d\Psi'\, g_J(\Psi) \,C_J^{-1}(\Psi,\Psi')\,g_J(\Psi')\\
&\hspace{2cm}+\sum_{k=1}^N\,\log\!\int d\Psi  e^{-iz\Psi^\dagger\Psi} P_{d_k}(g_1(\Psi),\ldots,g_{d_k}(\Psi); e^{2i \Psi^\dagger \Psi},e^{4i \Psi^\dagger \Psi},\ldots).\nn
\end{align}
What remains is to understand the behavior of this expression for large graphs, the saddle point structure that emerges, and to recognize the
considerable algebraic simplifications that occur in this regime.

\section{The large $N$ saddle point}\label{secsddl}

We have already assumed in (\ref{maxdeg}) that the maximal degree is $D$. Then, by rearranging the vertices, we can make the first $N_1$ vertices have degree 1, the subsequent $N_2$ vertices have degree 2, and so on. We will then take the limit $N\to\infty$ keeping the ratios
\beq\label{xddef}
x_d\equiv \frac{N_d}{N}
\eeq
fixed. We are not assuming that all the $x$'s are nonzero. In fact, the typical cases we consider will have a limited set of degrees present.

In the $N\to\infty$ limit characterized by the ratios (\ref{xddef}), our auxiliary field representation for the resolvent given by (\ref{resaux}-\ref{Seff}) can be restructured to reveal a saddle point exponential controlled by $N$:
\begin{align}\label{resauxx}
&\frac1N\langle\Tr[\mathbf{A} - z\mathbf{I}]^{-1}\rangle  =\frac{1}{\mathcal{N}} \int \left(\prod_{d=1}^D\mathcal{D}g_d\right)e^{-S[g_1,\ldots,g_D]}\\
&\hspace{2cm}\times\sum_{d=1}^D x_d\frac{\int d\Psi (2i\phi^2) e^{-iz\Psi^\dagger\Psi} P_{d}(g_1(\Psi),\ldots,g_{d}(\Psi); e^{2i \Psi^\dagger \Psi},e^{4i \Psi^\dagger \Psi},\ldots)}{\int d\Psi  e^{-iz\Psi^\dagger\Psi} P_{d}(g_1(\Psi),\ldots,g_{d}(\Psi); e^{2i \Psi^\dagger \Psi},e^{4i \Psi^\dagger \Psi},\ldots)},\nn
\end{align}
\begin{align}\label{Seffx}
&S[g_1,\ldots,g_D]\equiv-\sum_{d=1}^D\frac{1}{2d}\int d\Psi\,d\Psi'\, g_d(\Psi) \,C_d^{-1}(\Psi,\Psi')\,g_d(\Psi')\\
&\hspace{2cm}+N\sum_{d=1}^D\,x_d\log\!\int d\Psi  e^{-iz\Psi^\dagger\Psi} P_{d}(g_1(\Psi),\ldots,g_{d}(\Psi); e^{2i \Psi^\dagger \Psi},e^{4i \Psi^\dagger \Psi},\ldots).\nn
\end{align}
The last line features an explicit factor of $N$ that tends to $\infty$, while $S$ appears in the exponent in (\ref{resauxx}). This is a classic saddle point
structure that suggests looking for the extremal points with respect to $g_d(\Psi)$ that dominate the large $N$ limit of (\ref{resauxx}). There is a subtlety, however: if we only take the last line of (\ref{Seffx}) as our saddle point function, no saddle point will exist, most obviously, because $P_D$, as defined by (\ref{Pdef}), is always a linear function of $\al_D$, and hence of $g_D(\Psi)$ in the context of (\ref{Seffx}). For that reason, no extrema of the second line of (\ref{Seffx}) with respect to $g_D(\Psi)$ may exist. The actual saddle point is determined by the balance of both lines in (\ref{Seffx}). This situation, and the discussion that follows, directly parallels the structure of the saddle point that dominates regular graph counting, successfully developed in \cite{regcount}.

We thus look for a saddle point of the form $\de S[g]/\de g_d=0$ that would control the large $N$ limit of (\ref{resauxx}). Explicitly, functional differentiation of $S$ yields the following integral equations:
\beq\label{sddlraw}
g_d(\Psi)=Nd\sum_{d'=1}^D x_{d'}\frac{\int d\Psi' e^{2id \Psi^\dagger \Psi'}e^{-iz\Psi'^\dagger\Psi'} \frac{\del}{\del g_d}P_{d'}(g_1(\Psi'),\ldots, g_{d'}(\Psi'); e^{2i \Psi'^\dagger \Psi'},\ldots)}{\int d\Psi'  e^{-iz\Psi'^\dagger\Psi'} P_{d'}(g_1(\Psi'),\ldots ,g_{d'}(\Psi'); e^{2i \Psi'^\dagger \Psi'},\ldots)},
\eeq
remembering that $C$ is defined by (\ref{Cdef}). Because the second line of (\ref{Seffx}) does not have a saddle point by itself, and the saddle point is determined by the balance of the first line of (\ref{Seffx}), which does not depend explicitly on $N$, and the $N$-dependent second line, it is natural that the position of the saddle point is $N$-dependent, as evident from the presence of an explicit factor of $N$ in (\ref{sddlraw}). It is then crucial to identify the leading $N$-scalings of the functions $g_d$ in order to be able to proceed with the saddle point analysis.

As mentioned above, the situation closely parallels the considerations of \cite{regcount} for the counting of regular graphs. The polynomials $P_d$ are essentially the same. The only difference is that the structures appearing in the saddle point considerations of \cite{regcount} are now decorated with
extra dependences on and integrations over $\Psi$, which do not affect the multiplicative $N$-scalings of $g_d$. In \cite{regcount}, the leading $N$-scalings of the saddle point configuration have been identified and verified in detail by comparisons with other known asymptotic results.
The analog of those scalings in our present context would be
\beq\label{sddleN}
g_d\sim N^{1-d/2}.
\eeq
It is easy to verify, in view of the power counting rules within $P_d$ mentioned under (\ref{Pdef}), that, under these scalings, the leading powers of $N$ assigned to the left-hand side and right-hand side of (\ref{sddlraw}) are indeed consistent.\footnote{Let us run a quick argument:
Since $g_1$ is much greater than the other $g_d$, the leading contributions always come from the highest powers of $g_1$. For $\del P_{d'}/\del g_d$ in the numerator of (\ref{sddlraw}), this corresponds to the $g_1^{d'-d}g_{d}$ term in $P_{d'}$, which yields after differentiation
$g_1^{d'-d}\sim N^{(d'-d)/2}$. In the denominator of (\ref{sddlraw}), the leading power of $N$ comes from $g_1^{d'}\sim N^{d'/2}$. Dividing the two powers and multiplying by the explicit prefactor of $N$ in (\ref{sddlraw}) gives $g_d\sim N^{1-d/2}$, consistent with the ansatz (\ref{sddleN}).} We do not have an obvious way to prove that no other saddle points exist and that they do not dominate the actual statistical expectation values, but we make the practical choice of focusing on the saddle point with the $N$-scalings given by (\ref{sddleN}), finding its contribution to the eigenvalue density, and successfully verifying the result against other known analytic results in special cases, and against numerics.

Assuming (\ref{sddleN}), the polynomials $P_d$ simplify tremendously. Indeed, with (\ref{sddleN}), $g_1\sim\sqrt{N}$, $g_2\sim 1$ and all higher $g_d$ vanish at large $N$. Then, $P_d$ is dominated by the term with the highest power of $g_1$, or of $\al_1$ in terms of (\ref{Pdef}), which is easily extracted by differentiation as 
\beq\label{Pg1}
P_d\approx \frac{(-g_1)^d}{d!}.
\eeq
Since this only depends on $g_1$, we can obtain a self-contained equation for $g_1$ from (\ref{sddlraw}). We furthermore only need $g_1$, and not the higher $g_d$, to compute a saddle point estimate of (\ref{resauxx}) within this approximation, which becomes exact at $N\to\infty$. Introducing $g$ so that
\beq\label{g1N}
g_1=\sqrt{N} g,
\eeq
we obtain from (\ref{sddlraw})
\beq\label{sddlsimp}
g(\Psi)=\sum_{d=1}^D x_{d}\,d\,\frac{\int d\Psi' e^{2i \Psi^\dagger \Psi'}e^{-iz\Psi'^\dagger\Psi'}[g(\Psi')]^{d-1}}{\int d\Psi'  e^{-iz\Psi'^\dagger\Psi'} [g(\Psi')]^d},
\eeq
Note that $N$ has dropped out from this equation, which is one indication that the powers of $N$ assigned by (\ref{sddleN}) are consistent.

As is customary in the applications of the Fyodorov-Mirlin method, we assume that the saddle point solution respects the `superrotation' symmetry of the equation: those linear transformations of $\Psi$ that leave invariant the scalar product (\ref{scalarp}). Then,
\beq\label{gsup}
g(\Psi)=g(\Psi^\dagger\Psi).
\eeq
In that case, the denominator in (\ref{sddlsimp}) can be evaluated using the following formula \cite{sup2} valid for any function $F(\rho)$ decreasing at $\rho\to\infty$:
\beq\label{F0}
\int d\Psi F(\Psi^\dagger\Psi)=\pi F(0).
\eeq
We thereby get
\beq\label{gPsig0}
\pi g(\Psi^\dagger\Psi) g(0)=\int d\Psi' e^{2i \Psi^\dagger \Psi'}e^{-iz\Psi'^\dagger\Psi'} \del_y X\Big|_{y=g(\Psi'^\dagger\Psi')/g(0)},
\eeq
where we have introduced the generating function of the degree distribution $x_d$ given by
\beq\label{xdef}
X(y)\equiv \sum_{d=1}^D x_d\, y^d.
\eeq
We can inspect (\ref{gPsig0}) at $\Psi=0$, which gives\footnote{The other solution $g(0)=-\sqrt{c}$ is, of course, also possible, but it would have given an equivalent contribution. Since we are normalizing the end result exactly via an indirect argument below, adding up these two identical contributions plays no role. By contrast, similar extra saddles are important for detailed counting of regular graphs, as expounded in \cite{regcount}.}
\beq\label{g0c}
g(0)=\sqrt{c},
\eeq
where $c$ is the average degree:
\beq\label{meandeg}
c\equiv\sum_{d=1}^D x_d \,d.
\eeq
We then obtain the following simplified form of the saddle point equation:
\beq\label{sddlsuper}
g(\Psi^\dagger\Psi) =\frac1{\pi\sqrt{c}}\int d\Psi' e^{2i \Psi^\dagger \Psi'}e^{-iz\Psi'^\dagger\Psi'} \del_y X\big|_{y=g(\Psi'^\dagger\Psi')/\sqrt{c}}\,\,\,.
\eeq
While all of our derivations have technically assumed that the maximal degree $D$ is finite, the end result in principle makes sense without this restriction, at least for sufficiently rapidly decreasing $x_d$. (We mean specifically that, as long as the degree distribution decays sufficiently rapidly so that (\ref{xdef}) is well-defined for all $y$, we can still employ the corresponding $X$ in (\ref{sddlsuper}) producing a meaningful equation. We shall see below that, for \ER graphs, this recovers the correct integral equations of \cite{BR,MF,FM,Khetal}. If the degree distribution is heavy-tailed, $X$ may not necessarily exist everywhere, which would make this strategy fail. Heavy-tailed degree sequences are generally known to lead to many subtleties, even for simpler problems involving graph counting \cite{tail}.)

The formulas at this point display some affinity to the mean-field approaches to random graphs, where one assumes that the graph is tree-like and the neighborhoods have a simple structure, and prescribes a degree distribution. The cavity approach \cite{cavity, suscaetal} is much in this spirit, as are the considerations of \cite{SC,SM}; see also \cite{MK} for a treatment of random walks on random graphs in this sort of mean-field language. In our case,
we started with rigidly controlled vertex degrees and arrived in the $N\to\infty$ limit at a saddle point equation in terms of the degree distribution.

Once a solution to (\ref{sddlsuper}) has been found, one needs to evaluate the saddle point estimate of (\ref{resauxx}) and then the eigenvalue density
from (\ref{rhodef}). First of all, we have the second line of (\ref{resauxx}), where we simply have to substitute the found solution for $g$ via (\ref{g1N}) taking into account that the relevant part of $P_d$ is $(-g_1)^d\!/d!$, as given in (\ref{Pg1}). On top of that, there are in principle various determinants from the remaining 
Gaussian integrals arising via the saddle point expansion, but those must simply cancel the normalization factor $\mathcal{N}$ by the general principles
already mentioned in \cite{laplace,hmmr}. Indeed, in parallel with the computation we presented, we could have started with the average of 1 instead of
$\Tr[\mathbf{A} - z\mathbf{I}]^{-1}/N$ in (\ref{resav}), and proceeded with all of our derivations line-by-line arriving at (\ref{resauxx}-\ref{Seffx}) but without the second line of (\ref{resauxx}). The saddle point equation and all the Gaussian integration factors would have been unchanged, but the result must tautologically be 1 since we started with a probabilistic averaging of 1. This tells us that, for the average of $\Tr[\mathbf{A} - z\mathbf{I}]^{-1}/N$ in (\ref{resav}), which is what we are actually considering, the result must be exactly equal to the second line of (\ref{resauxx})
evaluated at (\ref{g1N}), with $g$ taken as the solution of (\ref{sddlsuper}). Then, by (\ref{rhodef}),
\begin{align}\label{psuper}
p(\la)&=\frac1{\pi^2}\Im\sum_{d=1}^D x_d\int d\Psi (2i\phi^2)\,e^{-iz\Psi^\dagger\Psi}  \left(\frac{g(\Psi^\dagger\Psi)}{g(0)}\right)^d\\
&\hspace{2cm}=\frac1{\pi^2}\Im\int d\Psi (2i\phi^2)\,e^{-iz\Psi^\dagger\Psi} 
X\big|_{y=g(\Psi^\dagger\Psi)/\sqrt{c}}\,\,\,,\nn
\end{align}
where we have used (\ref{F0}) to evaluate the denominators in (\ref{resauxx}) at the saddle point, and then (\ref{xdef}-\ref{g0c}).

What remains is to test the operation of (\ref{sddlsuper}) and (\ref{psuper}) in concrete cases.

\section{McKay distribution}\label{secMcKay}

Consider first the case of $D$-regular graphs, so that $x_{d<D}=0$, $x_D=1$ and
\beq
X(y)=y^D,\qquad c=D.
\eeq
Equations  (\ref{xdef}-\ref{g0c}) then turn into
\beq\label{sddlD}
g(\Psi^\dagger\Psi) =\frac1{\pi D^{(D-2)/2}}\int d\Psi' e^{2i \Psi^\dagger \Psi'}e^{-iz\Psi'^\dagger\Psi'} [g(\Psi'^\dagger\Psi')]^{D-1}
\eeq
and
\beq\label{psupD}
p(\la)=\frac1{\pi^2D^{D/2}}\Im\int d\Psi (2i\phi^2)\, e^{-iz\Psi^\dagger\Psi} [g(\Psi^\dagger\Psi)]^{D}.
\eeq
A key observation for solving (\ref{sddlD}) is that it respects the Gaussian ansatz $g=b\, e^{-a\Psi^\dagger\Psi}$. Indeed, if $g$ is of this form, we get a Gaussian integral on the right-hand side of (\ref{sddlD}), which consistently evaluates into a Gaussian function that matches the left-hand side under appropriate identifications of $a$ and $b$ as functions of $z$.

We can perform the above manipulations explicitly by writing
\begin{align*}
\int d\Psi' e^{2i \Psi^\dagger \Psi'}e^{-[(D-1)a+iz]\Psi'^\dagger\Psi'}=\int d\phi'd\chi'd\xi'd\eta'& e^{2i (\phi\phi'+\chi\chi')}(1+\cdots)\\
&\hspace{-1cm}\times e^{-[(D-1)a+iz](\phi'^2+\chi'^2)}[1-[(D-1)a+iz]\xi'\eta'],
\end{align*}
where `$\cdots$' denotes the terms dependent on $\xi$ and $\eta$, which could easily be computed directly with a bit of extra work, but which must
match on the two sides of (\ref{sddlD}) automatically by superrotation invariance as long as the $\xi\eta$-independent terms match  \cite{MF,FM,laplace}. Then, by the integration rules (\ref{Berezin}), we get
\begin{align*}
\int d\Psi' e^{2i \Psi^\dagger \Psi'}e^{-[(D-1)a+iz]\Psi'^\dagger\Psi'}&=[(D-1)a+iz]\int d\phi'd\chi'e^{2i (\phi\phi'+\chi\chi')}e^{-[(D-1)a+iz](\phi'^2+\chi'^2)}+\cdots
\\
&=\pi e^{-(\phi^2+\chi^2)/[(D-1)a+iz]}+\cdots
\end{align*}
At the same time, $g(\Psi^\dagger\Psi)=b\,e^{-a(\phi^2+\chi^2)}+\cdots$, so the matching of the two sides of (\ref{sddlD}) gives
\beq\label{eqab}
a=\frac1{(D-1)a+iz},\qquad b=\frac{b^{D-1}}{D^{(D-2)/2}}. 
\eeq
Then, $b=\sqrt{D}$ in accord with (\ref{g0c}), and
\beq
a=\frac{-iz+\sqrt{4(D-1)-z^2}}{2(D-1)}.
\eeq
To evaluate (\ref{psupD}), we need to consider
\begin{align*}
&\int d\Psi (2i\phi^2) e^{-(Da+iz)\Psi^\dagger\Psi}=\int d\phi\, d\chi\, d\xi\, d\eta\, (2i\phi^2)\, e^{-(Da+iz)(\phi^2+\chi^2)}[1-(Da+iz)\xi\eta]\\
&\hspace{3cm}=(Da+iz)\int d\phi\, d\chi (2i\phi^2)\, e^{-(Da+iz)(\phi^2+\chi^2)}=\frac{i\pi}{Da+iz}.
\end{align*}
Substituting this into (\ref{psupD}), we get:
\beq
p_{\mathrm{McKay}}(\la)=\frac1\pi\Re\frac{1}{Da+iz}\Bigg|_{z=\la}=\frac{\Re(D\bar a-i\bar z)}{\pi\,\,|Da+iz|^2}\Bigg|_{z=\la}=\frac{D\sqrt{4(D-1)-z^2}}{2\pi(D^2-z^2)}.
\eeq
This is the correct eigenvalue distribution for the adjacency matrices of large $D$-regular graphs, originally derived by McKay in \cite{mckay} using the method of moments. (Connections between this distribution and free probability theory have been discussed in \cite{mckayfree}.)

\section{Mixed-regular graphs}\label{secmix}

Having settled the simple case of $D$-regular graphs, we now return to the full formulation (\ref{sddlsuper}-\ref{psuper}) involving mixed degrees.
A common way to process equations like (\ref{sddlsuper}) is to introduce, in place of the commutative components of supervectors $\Psi$, polar coordinates of the form
\beq\label{polar}
\rho=\phi^2+\chi^2,\qquad \phi=\sqrt{\rho}\cos\alpha,\qquad \chi=\sqrt{\rho}\sin\alpha,\qquad d\phi\,d\chi=\frac12 d\rho\, d\alpha,
\eeq
so that
\beq\label{supersymm}
\Psi^\dagg\Psi=\rho+\xi\eta,\qquad g(\Psi^\dagg\Psi)=g(\rho)+\del_\rho g(\rho)\,\xi\eta,
\eeq
and then perform the integrations over the anticommuting components and the angle explicitly \cite{FM,MF,laplace}. One is then left with a one-dimensional integral equation for $g(\rho)$. In application to (\ref{sddlsuper}), this strategy yields
\begin{equation}
g(\rho) =- \frac{1}{\sqrt{c}}\int_0^\infty\hspace{-2mm} d\rho' J_0(2\sqrt{\rho\rho'})\,\del_{\rho'}\Big\{e^{-iz\rho'}\del_y X\big|_{y=g(\rho')/\sqrt{c}}\Big\},
\label{eq:saddle}
\end{equation}
where $X$ is the generating function of the degree distribution defined by (\ref{xdef}), $c$ is the average degree defined by (\ref{meandeg}), and $J_0$ is the standard Bessel function. Similarly, from (\ref{psuper}), we get
\beq\label{prho}
p(\la)=\frac{1}{\pi}\,\Re\!\int_0^\infty \hspace{-2mm}d\rho \,e^{-iz\rho}\, \,X\!\left(\frac{g(\rho,z)}{\sqrt{c}}\right)\Bigg|_{z=\la}\,\,\,.
\eeq
We can integrate by parts in (\ref{eq:saddle}) and obtain, introducing $G(\rho)\equiv g(\rho)/\sqrt{c}-1$,
\begin{equation}
G(\rho) =-\frac{\rho}{c} \int_0^\infty \hspace{-2mm} d\rho' \,\frac{J_1(2\sqrt{\rho\rho'})}{\sqrt{\rho\rho'}}\,e^{-iz\rho'}\del_y X\Big|_{y=1+G(\rho')}\,\,\,.
\label{Gsaddle}
\end{equation}

The structure of (\ref{Gsaddle}) is that of a nonlinear integral equation for $G(\rho)$, which also depends parametrically on $z$, but the equation
is solved independently at each $z$-point. The patterns involved go under the name of Hammerstein equations in the mathematical literature \cite{surv}. Such equations, specifically with Bessel kernels, appear frequently in considerations of sparse random graph problems \cite{BR,RB,MF,FM,laplace,walkongraph,gel,Khetal,inhmg}. We have now demonstrated how they emerge for the spectra of random graphs with prescribed degree sequences.

The derivations of section~\ref{secMcKay} can, of course, be reproduced in the language of (\ref{eq:saddle}) or (\ref{Gsaddle}). Indeed, if the graph is $D$-regular and $X=y^D$, the ansatz $g=b\hspace{0.5mm}e^{-a\rho}$ is consistent, and the integral equation reduces to algebraic equations for $a$ and $b$, whereupon
the McKay distribution is recovered from (\ref{prho}). If the graph is not regular and multiple degree values are present, this approach evidently fails, since plugging in a single exponential for $g$ on the right-hand side of (\ref{eq:saddle}) returns $g$ as a sum of multiple exponentials. Repeating this process iteratively leads to infinite proliferation of exponentials, and this picture has been used in \cite{gel} to develop numerical approximations to solutions
of similar integral equations. It is also closely related to the considerations in \cite{kuehn}. We shall, however, opt for more down-to-earth methods
in our numerical analysis of (\ref{eq:saddle}) below.

A curious further check of our formalism is provided by taking the degree distribution to be Poissonian with mean $c$, that is, $x_k=c^ke^{-c}/k!$, and hence
$X(y)=e^{c(y-1)}$. This case corresponds to sparse \ER graphs frequently studied in the literature. Indeed, substituting $\del_y X=c e^{c(y-1)}$ into
(\ref{Gsaddle}) results, after the redefinition $cG(\rho)\to ig(\rho)$, in the specific Bessel-kernel Hammerstein equation with an exponential nonlinearity seen in the literature  for \ER graphs
\cite{BR,FM,MF,Khetal,hmmr}.

Equations of the form (\ref{Gsaddle}) can be effectively solved numerically. The general principles are outlined in \cite{surv}, while we have recently
displayed in \cite{hmmr} a simple and successful implementation specifically for Bessel kernels, focusing on equations arising from
sparse \ER graph problems. The general principle is that one expands the unknown in a finite basis of `test functions' $\ph_j$ and inserts an appropriate projector in the equation so that solutions exist within the subspace spanned by $\ph_j$. In such a formulation, (\ref{Gsaddle}) turns,
approximately, into a system of nonlinear equations for the expansion coefficients associated with $\ph_j$. There are different choices possible
for the space of $\ph_j$ and the projector on this space necessary for implementing the above construction, and one must choose wisely to ensure effective convergence to true solutions of (\ref{Gsaddle}).

In \cite{hmmr}, we focused on collocation methods, and specifically on their variation proposed in \cite{KS}, where one expands the full nonlinearity
$\del_y X$ in (\ref{Gsaddle}), rather than the unknown $G$ itself, in the chosen basis $\ph_j$. This has the advantage of minimizing the number of integrals that actually have to be computed, since the expansion of $\del_y X$ in terms of $\ph_j$ appears inside the integral linearly. This method has worked very well in \cite{hmmr}, but in our present setting we find it less advantageous, since at the end of the day, one still has to recover
the eigenvalue density by performing the integral in (\ref{prho}). For the case of \ER graphs treated in \cite{hmmr}, $X(y)=e^{c(y-1)}$, so $X$ can be linearly expressed through $\del_y X$. With this feature, we could implement all the integrals in \cite{hmmr} analytically using a Laguerre basis.
This is no longer possible in our present setting, since in general a linear expansion of $\del_y X$ in the prescribed basis $\ph_j$ does not produce any simple nice formula for $X$ that can be integrated analytically in (\ref{prho}), so at least some numerical integrations will be needed if implementing the strategy of \cite{KS} following the considerations of \cite{hmmr}, and hence we find this approach suboptimal.

Instead, we propose here an alternative scheme for numerically approximating the solutions of (\ref{Gsaddle}) that will only involve analytic
integrations. Our main target is graphs where the maximal degree $D$ is finite, and the possible degrees $d=2..D$ are present\footnote{One could also include vertices of degree 1, but we omit them as they are not particularly interesting and would require writing extra terms in the formulas that are structured somewhat differently from the contributions with $d\ge 2$.} in fixed proportions
$x_d$ as the size of the graph $N$ is sent to infinity, which may be referred to as `mixed-regular graphs.' In this case, $X$ is a degree $D$ polynomial in $y$, and $\del_y X$ is a degree $D-1$ polynomial. We find it convenient to go back to (\ref{eq:saddle}) for our implementation,
and expand $g$ through test functions $\ph_j$ chosen as products of Laguerre polynomials $L_j$ and exponentials:
\beq\label{gexp}
g(\rho)=Q(\ga\rho) e^{-\ga\rho},\qquad Q(x)\equiv\sum_{j=0}^J \beta_jL_j(x),\qquad x\equiv \ga\rho.
\eeq
Note that $Q$ is a general degree $J$ polynomial expressed through its Laguerre basis components $\be_j$.
Our choice of the Laguerre basis is motivated by the integrals extending from 0 to $\infty$ in (\ref{eq:saddle}).
We will never use the orthogonality of Laguerre polynomials on the half-line explicitly, but they provide a good expansion basis
and are known to work effectively, as discussed in the mathematical literature and our recent article \cite{hmmr}.
We treat $\ga$ and $J$ as adjustable parameters that control the performance of our numerical approximation scheme.

To evaluate the integrals in (\ref{eq:saddle}), we use the following expansion of the Bessel function \cite{Sz}:
\beq
J_0(2\sqrt{xy}/\ga)=e^{-x}\sum_{n=0}^\infty \frac{x^n\,L_n(y/\ga^2)}{n!},
\eeq
where $y/\ga$ is identified with $\rho'$ in (\ref{eq:saddle}). We have also observed that the numerical convergence is more reliable if 
we do not replace $g(0)$ with $\sqrt{c}$ following (\ref{g0c}), but rather keep it as $g(0)$ and let it track the convergence process. Thus,
where we see $\sqrt{c}$ in (\ref{eq:saddle}-\ref{prho}), we must restore it as $g(0)$. Furthermore, from (\ref{gexp}),
\beq\label{g0be}
g(0)=\sum_{j=0}^J\be_j,
\eeq
which will of course converge back to the value $\sqrt{c}$ on solutions.
Put together, this yields in place of (\ref{eq:saddle})
\begin{align}
\sum_{j=0}^J \be_j  L_j(x)=&- \mathbb{P}\sum_{n=0}^\infty\frac{x^n}{n!}\int dy\,
 L_n(y/\ga^2)\,e^{-izy/\ga}\\
&\times\sum_{d=2}^D \frac{ x_d\,d(d-1)e^{-(d-1)y}}{(\sum \be_j)\rule{0mm}{3.5mm}^{d}}\left[\del_{y}Q(y)-\left(1+\frac{iz}{\ga(d-1)}\right)Q(y)\right]Q^{d-2}(y).\nn
\end{align}
Here, $\mathbb{P}$ denotes a projector on the space of degree $J$ polynomials that we still have to choose. A very natural choice
adapted to our present analytic formulas
(though distinct from what is typically used in generic expositions of numerical methods for Hammerstein equations) is
to simply truncate the $n$-sum on the right-hand side at $n=J$.
We then get:
\begin{align}\label{eqJ}
\sum_{j=0}^J \be_j  L_j(x)=&- \sum_{n=0}^J\frac{x^n}{n!}\int dy\,
 L_n(y/\ga^2)\,e^{-izy/\ga}\\
&\times\sum_{d=2}^D \frac{ x_d\,d(d-1)e^{-(d-1)y}}{(\sum \be_j)\rule{0mm}{3.5mm}^{d}}\left[\del_{y}Q(y)-\left(1+\frac{iz}{\ga(d-1)}\right)Q(y)\right]Q^{d-2}(y).\nn
\end{align}
The two sides are manifestly polynomials of degree $J$ in $x$, and hence equating the $J+1$ independent coefficients provides $J+1$ nonlinear equations for the $J+1$ unknowns $\be_j$. Once $\be_j$ have been found, $g$ is reconstructed from (\ref{gexp}), and the eigenvalue density given by (\ref{prho}) is reconstructed from
\beq\label{pmixreg}
p(\la)=\frac{1}{\pi\gamma}\Re\!\int_0^\infty \hspace{-2mm}dx \,e^{-izx/\ga} \sum_{d=2}^D\frac{ x_d\,e^{-dx}}{(\sum \be_j)\rule{0mm}{3.5mm}^{d}}\,Q^{d}(x)\Big|_{z=\la}\,\,,
\eeq
again, with $\sqrt{c}$ in (\ref{prho}) equivalently replaced with $g(0)$ given by (\ref{g0be}), as explained above.
Note that all the integrands in (\ref{eqJ}) and (\ref{pmixreg}) consist of polynomials and exponentials. If the polynomials are appropriately multiplied and expanded out through monomials, which is easily implemented using computer algebra, each monomial is integrated analytically by applying
\beq
 \int_0^\infty \hspace{-2mm}dy\, y^m\,e^{-\al y} =\frac{m!}{\al^{m+1}}.
\eeq
There are no numerical integrations involved.

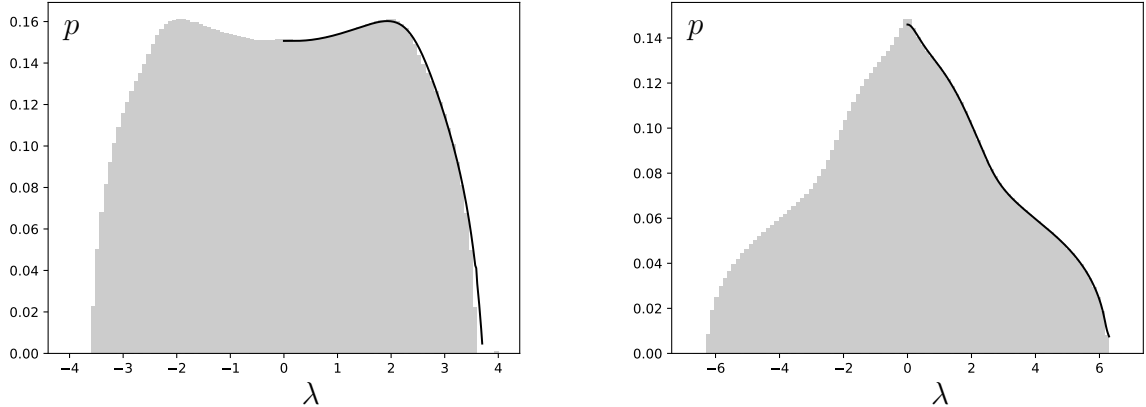
\begin{figure}[t]\vspace{-10mm}
\centering
\includegraphics[width = 0.49\linewidth]{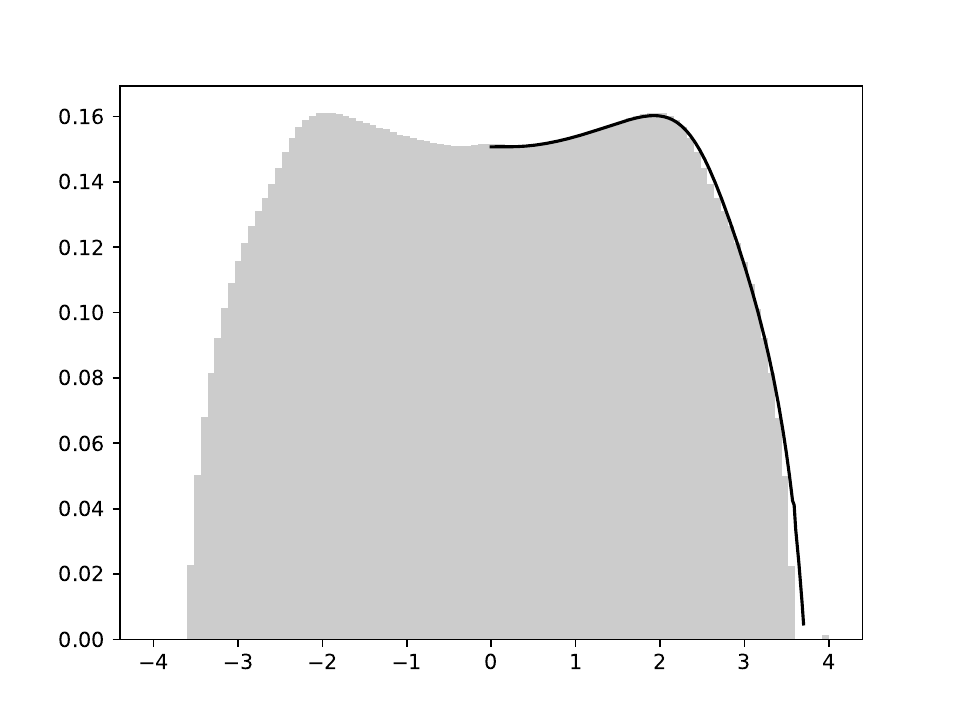}\hspace{2mm}\includegraphics[width = 0.49\linewidth]{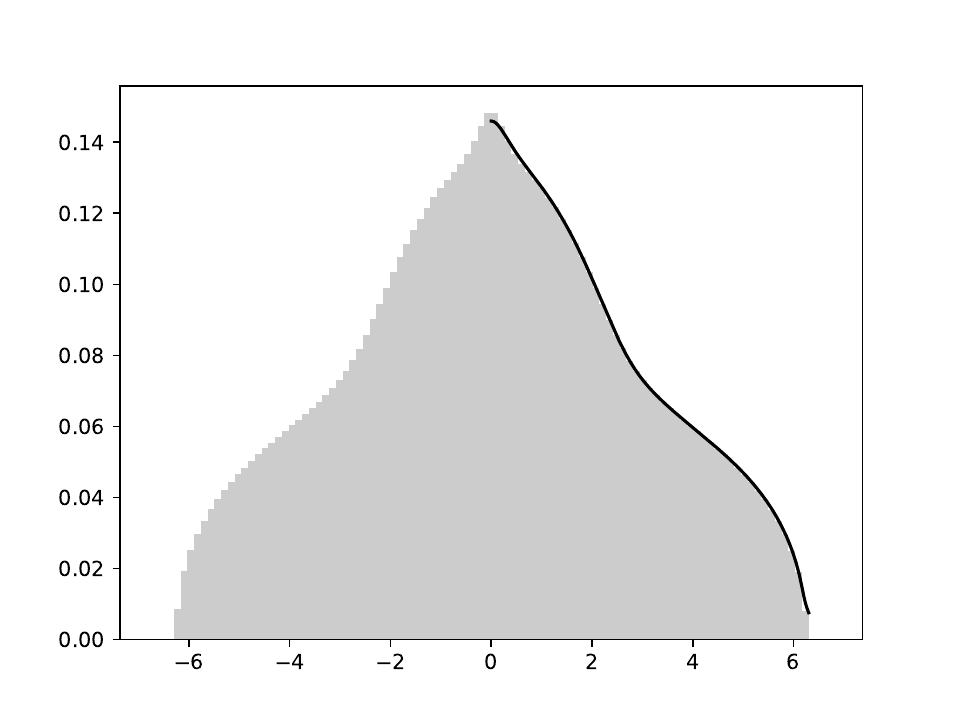}
\begin{picture}(0,0)
\put(-435,140){$p$}
\put(-345,0){$\la$}
\put(-199,140){$p$}
\put(-107,0){$\la$}
\end{picture}\vspace{2mm}
\caption{Solutions of the nonlinear system (\ref{eqJ}) for $\be_j$, followed by reconstructing the eigenvalue densities (\ref{pmixreg}), plotted as solid black lines {\bf (left)} for $70\%$ of degree 3 vertices and $30\%$ of degree 5 vertices, solved at $J=7$, $\ga=2$ and {\bf (right)} for $50\%$ of degree 4 vertices and $50\%$ of degree 12 vertices, solved at $J=7$, $\ga=3/2$. As the distributions are reflection-symmetric, only the $\la>0$ parts of the analytic predictions (solid black lines) are plotted explicitly. For the initial seed in the root search algorithm, $\be_0=1$ and $\be_{j>0}=0$ is used at $\la=0$, and the previous solution is reused as the seed for each next value of $\la$. The grey shaded areas represent empirical eigenvalue density histograms obtained from a sample of 500 random graphs with 10000 vertices each and the corresponding degree proportions.}
\label{fig}
\end{figure}
We have created a basic implementation of the above process in Python (a sample script is provided in the Appendix) using the {\tt root}
function of the SciPy library \cite{SciPy} to search for the roots of the nonlinear algebraic system (\ref{eqJ}). We have then compared the output
(taking for concreteness graphs with two possible degree values occurring in a prescribed proportion) with the empirical histograms obtained
by random sampling. To generate graphs with prescribed degree sequences, we used the {\tt Degree\_Sequence} function of the igraph library \cite{igraph} with the `{\tt edge\_switching\_simple}' option. (For algorithmic aspects of generating random graphs with prescribed degrees, 
see \cite{degseqgen}.) The results are displayed in Fig.~\ref{fig} and they show excellent agreement between solving (\ref{eq:saddle}) numerically, followed by reconstructing the eigenvalue density from (\ref{prho}), and empirical histograms.

\section{Conclusions}\label{secconc}

We have developed a statistical field theory treatment of the eigenvalue distributions of adjacency matrices corresponding to random graphs
with prescribed degrees. The final output of our theory is in the form of the nonlinear integral equation (\ref{Gsaddle}), where the nonlinearity is defined by $X$, which is the generating function of the degree distribution. Given solutions of such equations, the eigenvalue density can be recovered
by applying (\ref{prho}).

A number of steps in our derivations are admittedly heuristic, though firmly rooted 
in the statistical physics lore: the usage of functional integrals and functional saddle points in the Fyodorov-Mirlin approach to random matrix eigenvalue distributions, and the self-consistent assumptions about the dominant saddle point: the superrotation invariance (\ref{gsup}) and the leading large $N$ scalings (\ref{sddleN}). Despite these heuristic aspects, we can be confident about the end results, since (\ref{Gsaddle}) passes a number of stringent tests. First, for $D$-regular graphs, one immediately recovers the known McKay distribution. Curiously, this distribution emerges from the solution 
of elementary algebraic equations (\ref{eqab}), while all the highly nontrivial combinatorics of graph path counting, typical of conventional derivations based on the method of moments \cite{mckay}, is effectively stored inside compact and explicit closed-form functions, with no numerical coefficients more complicated than 2 or $\pi$ involved. Second, if we substitute the Poissonian degree distribution of sparse \ER graphs, we correctly recover the integral equation that controls the eigenvalue spectra of such graphs. This equation has been previously derived in a number of ways \cite{BR,FM,MF},
including the method of moments \cite{Khetal}, which is mathematically rigorous, even if laborious computationally. (The equations we derive also appear to be closely related to the content of Theorem 2 in \cite{reslarge}, though expressed in a vastly different language.) Finally, we have developed numerical solutions of the nonlinear integral equations that correctly reproduce the eigenvalue distributions for the mixed-degree case, some of them with
rather ornate, statuesque shapes, as in Fig.~\ref{fig}.

Spectra of adjacency matrices of random graphs with prescribed degrees have been previously studied at different levels of approximation using a variety of methods that go under the names of `cavity,' `population dynamics' and `message passing,' see \cite{empi,spectop,message}. Our present contribution complements these considerations by deriving
a compact one-dimensional nonlinear integral equation (\ref{Gsaddle}) which can be solved using fully deterministic numerical methods, whereafter the eigenvalue density is recovered straightforwardly. For comparison, the analytic derivations in the well-known work \cite{spectop} lead to multidimensional integral equations
that are treated numerically by stochastic numerical simulations. One-dimensional equations similar to (\ref{Gsaddle}) can be seen in the literature in relation to the much simpler case of \ER graphs, starting with \cite{BR}. We find it striking that, when moving on to the much more complicated case of graphs with prescribed degrees, the only change is that the nonlinearity in the corresponding equation is altered, and the exponential nonlinearity typical of \ER graphs gets replaced with a more general nonlinearity constructed from the generating function of the degree distribution.

We have provided a basic numerical implementation in our treatment for reconstructing the eigenvalue density by solving the nonlinear integral equations, but there is considerable room for improvement in terms of pushing it to arbitrarily high precision and in stabilizing the convergence near the edges and sharp features of the eigenvalue distribution. We hope that systematic mathematical work on numerical methods for Hammerstein equations will provide improved techniques for solving (\ref{Gsaddle}).

We have phrased our considerations for adjacency matrices, leaving aside for concreteness the topic of graph Laplacians. For regular graphs,
the spectra of adjacency matrices and Laplacians are trivially related, but this is not the case once mixed degrees have been introduced. It should be straightforward
to develop a similar treatment for graph Laplacians by incorporating the supervector representations for the corresponding resolvents as used in \cite{laplace} and
the approach to implementing the degree constraints developed in this paper.

Anderson localization on graphs has received considerable attention \cite{loc1,loc2,loc3,loc4}, including the case of random graphs with prescribed degrees \cite{locdeg}. We hope that the techniques we have introduced here will contribute to furthering the analytic understanding of those models.


\section*{Acknowledgments}

We thank Roland Speicher for a discussion on treatments of random regular graphs in the language of free probability theory, and
Pierfrancesco Dionigi for drawing our attention to related mathematical works.
OE is supported by Thailand NSRF via PMU-B, grant number B13F670063. We acknowledge the use of computational resources of LPTMS at Universit\'e Paris-Saclay for generating and diagonalizing large samples of random matrices.\vspace{5mm}

\appendix

\section*{Appendix: \parbox[t]{10cm}{Python code for eigenvalue density of\\ mixed-regular graphs}}

We provide a basic Python script that solves (\ref{eq:saddle}) and computes the eigenvalue density from (\ref{prho}). We focus on graphs with $N_1$ vertices of degree $d_1$ and $N_2$ vertices of degree $d_2$, with $x_{1,2}\equiv N_{1,2}/N$. The script can be straightforwardly adapted to more complicated degree mixtures.

\begin{verbatim}
import numpy as np
import numpy.polynomial as pl
from scipy.optimize import root
import igraph as ig
import matplotlib.pyplot as plt

def eqs(beta,z):
 betac=beta[:J]+1j*beta[J:]
 betasum=sum(betac)
 coeff1=coef1/betasum**d1
 coeff2=coef2/betasum**d2
 Q=pl.Polynomial(pl.laguerre.lag2poly(betac))
 dQ=Q.deriv()
 Q1=coeff1*(dQ-Q*(1+1j*z/gamma/(d1-1)))*Q**(d1-2)
 Q2=coeff2*(dQ-Q*(1+1j*z/gamma/(d2-1)))*Q**(d2-2)
 rhs=1j*np.zeros((J))
 for n in range(J):
  Ln=pl.laguerre.lag2poly([0]*n+[1])
  for j in range(n+1):
   Ln[j]*=1/gamma**(2*j)
  Lnpoly=pl.Polynomial(Ln)
  polint1=(Lnpoly*Q1).coef
  denom=fac[n]
  for m in range(len(polint1)):
   denom*=d1-1+1j*z/gamma
   rhs[n]+=fac[m]*polint1[m]/denom
  polint2=(Lnpoly*Q2).coef
  denom=fac[n]
  for m in range(len(polint2)):
   denom*=d2-1+1j*z/gamma
   rhs[n]+=fac[m]*polint2[m]/denom
 eqs=betac+pl.laguerre.poly2lag(rhs)
 return np.append(np.real(eqs),np.imag(eqs))
 
def get_p(beta,z):
 betac=beta[:J]+1j*beta[J:]
 betasum=sum(betac)
 Q=pl.Polynomial(pl.laguerre.lag2poly(betac))
 Q1=(x1*(Q/betasum)**d1).coef
 Q2=(x2*(Q/betasum)**d2).coef
 out=0
 denom=1
 for m in range(len(Q1)):
  denom*=d1+1j*z/gamma
  out+=fac[m]*Q1[m]/denom
 denom=1
 for m in range(len(Q2)):
  denom*=d2+1j*z/gamma
  out+=fac[m]*Q2[m]/denom
 return np.real(out)/(gamma*np.pi) 
 
d1=4
d2=12
x1=0.5
x2=0.5
coef1=x1*d1*(d1-1)
coef2=x2*d2*(d2-1)
J=8 #corresponds to J+1 in the notation of the paper
gamma=1.5
zs=np.linspace(0,6.4,100)

fac=np.ones((max(d1,d2)*J+1))
for i in range(1,max(d1,d2)*J):
 fac[i+1]=(i+1)*fac[i]

p=[]   
beta=np.array([1]+[0]*(2*J-1))
for z in zs:
 sol=root(eqs,beta,args=(z),tol=1e-12)
 beta=sol.x
 p.append(get_p(beta,z)) 

plt.plot(zs,p,'k')
plt.ylim(0)
plt.show()

\end{verbatim}


\end{document}